# New Equation for Bending Development of Arbitrary Rods and Application to Palm Fronds Bending

Mikrajuddin Abdullah

Department of Physics, Bandung Institute of Technology,

Jalan Ganeca 10 Bandung 40132 Indonesia.

Tel. +62-22-2500834/Fax. +62-22-2506452

and

Bandung Innovation Center

Jalan Sembrani 20 Bandung, Indonesia

Email: mikrajuddin@gmail.com

**Abstract**

A new general equation to explain bending of arbitrary rods (from arbitrary materials, cross sections, densities, strengthnesses, bending angles, etc) was proposed. This equation can solve several problems found in classical equations, which have many limitations such as only applies for small bending angles or must be solved using very complex schemes. Experiments were also conducted to confirm the theroretical predictions. The equation might be used to explain bending of palm fronds in a very simple way. The proposed equation may be used to obtain solution of several problems which are usually btain with iteration procedures.

# 1. INTRODUCTION

One interesting phenomenon in nature is bending of palm fronds, banana leaves, tree trunks, etc. Such bendings are highly dependent on frond or trunk size, age, strengthness, number of leaves attached, etc. Indeed the problem of solid bendings can be solved using equations that have been developed long time ago in engineering. One famous equation is the Euler-Bernoulli equation, applied to bendings with small slopes [Stephen, 2007]. The Euler-Bernoulli equation is simple enough, so that in many cases, the analytic solutions can be obtained easily. However, the bending angle of the palm fronds, banana leaves, or tree trunks is generally large. **Figure 1** is pictures of some objects in natures which experience bending at large bending angle. In such cases, the Euler-Bernoulli equation can not be applied.

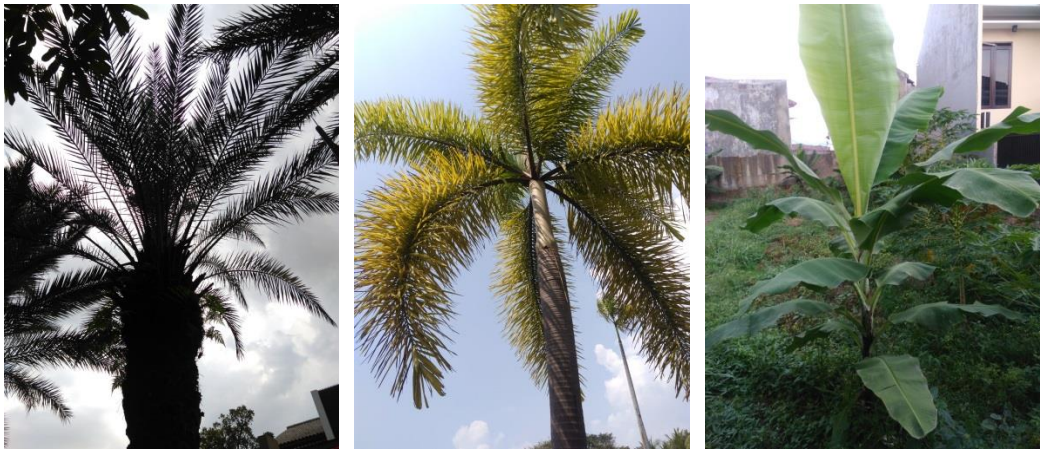

**Figurure 1** Pictures of bending of some palm fronds.

To best of our knowledge, discussions about bendings of natural objecs as shown in **Fig. 1** have not been reported, although these phenomena are very interesting to be explored. We suspect that the bending profiles of these objects are also related to stability under outside disturbances like windstorms. In all trees, the branch contributes a dynamic damping, which acts to reduce dangerous sway motion of the trunk and so minimizes loads and increases the mechanical stability [James, et al, 2006]. Same function should apply to palm fronds or banana leaves. Wind is the most persistent of the harmful natural forces to which any individual tree or forest stand is subjected [Jacobs, 1936].

Indeed, a quantitative structural analysis of a tree was attempted by Leonard Euler and later Greenhill in 1881. Both used static analyses to calculate the maximum critical height of a tree, above which it failed under its own weight [Spatz, 2000]. A tapered pole made of a homogeneous material was used as a simple model to approximate a tree. Modifications to the simple tree model include representing the canopy as a lumped mass on a column [Saunderson, et al, 1999], representing the tree as two masses (one for the canopy and one for the root-soil system) and the trunk as a weightless elastic column [Baker, 1995], and modeling the tree as a series of n logs with lumped masses representing branch whorls along the trunk [Guitard, et al. 1995]. However, detail discussion about the bending of palm fronds, banana leaves etc have not been reported.

These objects also have interesting shapes, the frond is attached by leaves of nearly the same spacing. If the distance between the leaves approaches zero, the palm fronds approaches the banana leaves. We observe that the banana leaf forms a bending profile like a palm fronds.

In this paper we derived a new equation for describing bending profile of any rods, which also applied to describing bending of palm fronds, banana leaves, and other objects in nature. The equation applies to any rod materials, cross sections, mass densities, strengthnesses, bending angles, etc. The classical problem of deflection of a cantilever beam of linear elastic material, under the action of a uniformly distributed load along its length (its own weight) and an external vertical concentrated load at the free end have been experimentally and numerically analysed by Beleândez et al [Beleândez, et al, 2003;2003]. The study was conducted on a long, thin, cantilever beam of uniform rectangular cross section made of a liniear elastic material that is homogeneous andisotropous.The beam wa assumed to be non-extensible and strais remain small, the plane cross-sections that are perpendicular to the neutral axis befor deformation remain plane and perpendicular to the neutral axis after deformation. The plane cross-section do no change their shape and area. They presented differential equation governing the behaviour of this system and show that this equation is rather difficult to solve due to the presence of a nonlinear term. Nevertheless, unless small deflections are considered, an analytical solution does not exist, since for large deflections a differential equation with a nonlinear term must be solved. The analysis of large deflection of cantilever beams of elastic materials has been discussed by Landau [Landau and Lifshitz, 1986] and the solution appears as elliptic integrals [Bisshopp and Drucker, 1945]

The derived equation was compared with observational data for bare rods as well as rods attached with pieces of plastics to duplicate the palm frond leaves. The derived equation is more easily solved rather than the well known differential equations. Indeed palms contain several thousand types of species [Bailey, 1971], however, in this work we only observe palms around our residence such as coconuts and palms in playgrounds.

## 2. METHOD

**Figure 2** illustrates bending of a rod of length $L$ positioned at fixed angle (formed by the fixed end to the horizontal direction) $\theta_0$. Let us divide the rod into $N$ segments having the same length, $a = L/N$. Consider the second segment which attracted by weight of rod section to the right having a length $L - 2a$, $W_1$, and non-gravitational forces acting at a distance $s_P$ from the fixed end. Both forces cause the segment to bend downward relative to the left one. The second segment has a slope $\theta_1$, smaller than $\theta_0$. The change in slope of the second segment to the first segment is $\Delta\theta_1 = \theta_0 - \theta_1$.

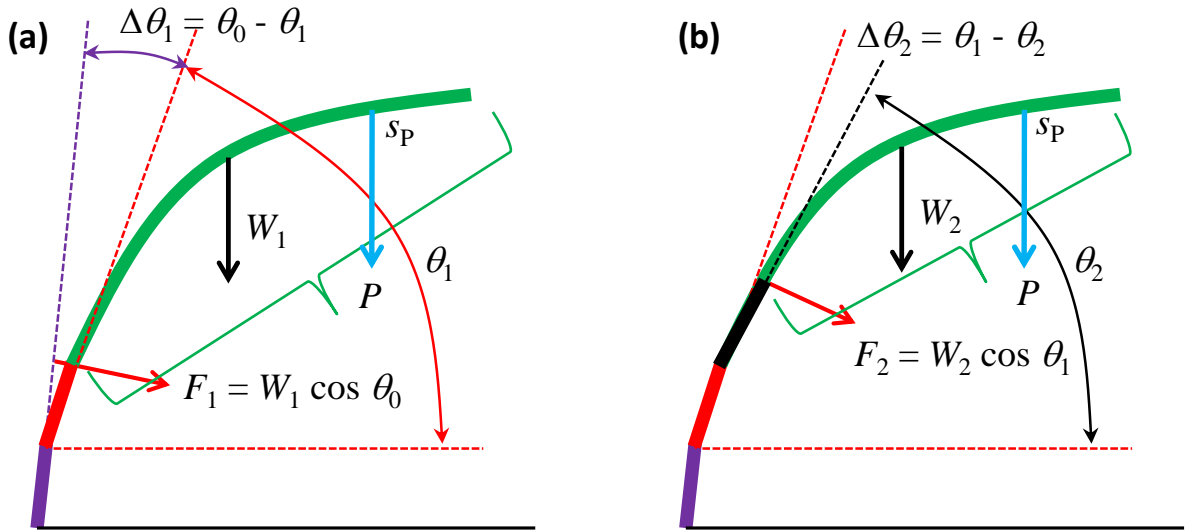

**Figure 2** Illustration of bending mechanism of the two leftmost segments of the rod.

The forces contributing to segment bending of the second segment are components that perpendicular to first segment direction, i.e. $F_1 = W_1 \cos\theta_0$ and $P\{s_P > a_1\}\cos\theta_0$. It should be noted that the initial direction of the second segment (prior to bending) is the same as the direction of the first segment. The notation of $P\{s_P > a_1\}$ means that this force is taken into account only if $s_P > a_1$. We hypotesize a change in the bending angle is proportional to force per unit area acting on the segment end as well as the segment length, or $\Delta\theta_1 = \alpha_1[(W_1\cos\theta_0 + P\{s_P > a_1\}\cos\theta_0)/A_1]a$, with $\alpha_1$ is the proportional constant and $A_1$ is the cross section of the second segment. We have applied similar hyphotesis when describing the bending of sparkling fireworks [Abdullah, et al 2014]. In general, the proportional constant depends on position along the rod. It might depend on the rod shape and size. The stronger the rod segment, the smaller the bending angle, indicating the smaller proportional constant. The proportional constant therefore can be considered to be inversely proportional to the strengthness of the material.

If $\lambda_j$ is mass per unit length of the $j$-th segment ($j$ = 1, 2, …, $N$) then $W_1 = \sum_{j=2}^{N}(\lambda_j a)g = ag\sum_{j=2}^{N}\lambda_j$ and $F_1 = ag\sum_{j=2}^{N}\lambda_j \cos\theta_0$. The bending angle of the second segment can be written as $\theta_1 = \theta_0 - (\alpha_1 a/A_1)\left(ga\sum_{j=2}^{N}\lambda_j + P\{s_P > a_1\}\right)\cos\theta_0$. The same analysis can be continued and we obtain an iteration equation for bending angle as follows

$$\theta_n = \theta_{n-1} - \frac{\alpha_n a}{A_n}\left(ga\sum_{j=n+1}^{N}\lambda_j + P\{s_P > na\}\right)\cos\theta_{n-1} \qquad (1)$$

Equation (1) is a general equation that applies to all types of rods, which can be rewritten as $(\theta_n - \theta_{n-1})/a = -(\alpha_n/A_n)\left(g\sum_{j=n+1}^{N}\lambda_j a + P\{s_p > na\}\right)\cos\theta_{n-1}$. We can change further this expression into differential form as following. Let us transform $\theta_{n-1} = \theta(s)$, $\theta_n = \theta(s+ds)$, $\alpha_n \approx \alpha_{n-1} = \alpha(s)$, $A_n \approx A_{n-1} = A(s)$, $a = ds$, $\sum_{j=n+1}^{N}\lambda_j a = \int_s^L \lambda(s')ds'$, and $P\{s_P > na\} = P\{s,L\}$. Based on this transformation, equation (1) becomes

$$\frac{d\theta}{ds} = -\frac{\alpha(s)\cos\theta(s)}{A(s)}\left\{\int_s^L g\lambda(s')ds' + P\{s,L\}\right\} \qquad (2)$$

The solution for equation (2) can be obtained by direct integration to result

$$\ln\left[\frac{\tan\theta + \sec\theta}{\tan\theta_0 + \sec\theta_0}\right] = -\int_0^s \frac{\alpha(z)}{A(z)}\left\{\int_z^L g\lambda(s')ds' + P\{z,L\}\right\}dz \qquad (3a)$$

or

$$\tan\theta + \sec\theta = [\tan\theta_0 + \sec\theta_0]\exp\left[-\int_0^s \frac{\alpha(z)}{A(z)}\left\{\int_z^L g\lambda(s')ds' + P\{z,L\}\right\}dz\right] \qquad (3b)$$

The solution for $\theta$ is

$$\theta = 2\tan^{-1}\left[\frac{\Omega-1}{\Omega+1}\right] \qquad (4)$$

with

$$\Omega = [\tan\theta_0 + \sec\theta_0]\exp\left[-\int_0^s \frac{\alpha(z)}{A(z)}\left\{\int_z^L g\lambda(s')ds' + P\{z,L\}\right\}dz\right] \qquad (5)$$

Equations (4) and (5) are general solution for bending angle for arbitrary rods. Equations (4) and (5) apply to both small and large bending angles.

*Palm Frond.* Generally, the fixed (attached) end of a palm frond has larger cross section than the free end. The free end grow later while the fixed end grow earlier, implies the frond part located near the fixed end has larger mechanical strength, and the mechanical strength decreases when moving to the free end. The proportional parameter then becomes higher when moving to the free end.

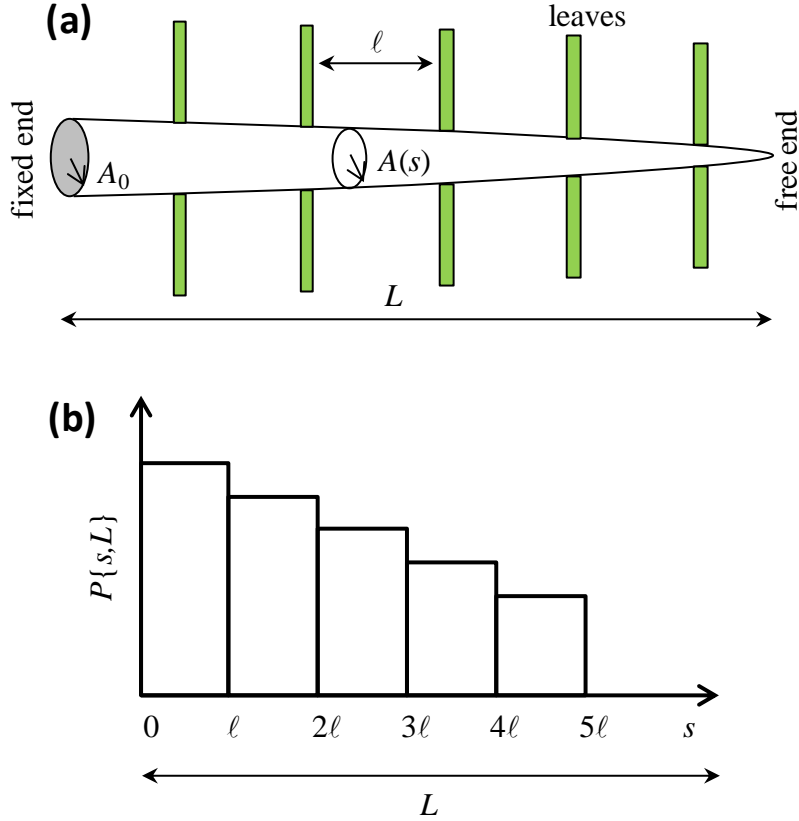

**Figure 3(a)** Illustration of palm frond shape with a cross secion is depending on the position. The left end is the fixed end attached to the tree and the right end is the free end. On a palm frond, the leaves are attached at the same spacing. **(b)** The shape of force $P\{s,L\}$ contributed by palm leaves. The leaves separation is $\ell$.

The cross sections of the palm frond at the fixed end and at a distance $s$ from the fixed end are $A_0$ and $A(s)$, respectively (**Fig. 3a**). The mass of the frond segment between $s$ and $s+ds$ is $dm = \mu A(s)ds$ with $\mu$ is the frond density per unit volume, and at present we assume to be constant. The mass of the frond per unit length at a distance $s$ from the fixed end becomes $\lambda(s) = dm/ds = \mu A(s)$. The leaves attached to the frond are considered as external forces acting on the frond. **Figure 3(b)** is a non-gravitational force acting on the frond. The force shape is a steplike function, which can be written as

$$P\{s,L\} = \begin{cases} Mmg & \text{if} & 0 < s \leq \ell \\ (M-1)mg & \text{if} & \ell < s \leq 2\ell \\ (M-2)mg & \text{if} & 2\ell < s \leq 3\ell \\ \vdots & \vdots & \vdots \\ mg & \text{if} & (M-1)\ell < s \leq M\ell = L \end{cases} \qquad (6)$$

with $M$ is the number of leaf positions, $m$ is the mass of each leaf, $q = s$ div $\ell$ (the lower round of $s/\ell$). Based on this shape we can write

$$\int_0^s \frac{\alpha(z)}{A(z)} P\{z,L\} dz = \left( \int_0^\ell + \int_\ell^{2\ell} + \int_{2\ell}^{3\ell} + \ldots + \int_{(q-1)\ell}^{q\ell} + \int_{q\ell}^s \right) \frac{\alpha(z)}{A(z)} P\{z,L\} dz \qquad (7)$$

Substituting $P\{s,L\}$ from Eq. (6) into Eq. (7) we have

$$\int_0^s \frac{\alpha(z)}{A(z)} P\{z,L\} dz = \left( \int_0^\ell Mmg + \int_\ell^{2\ell} (M-1)mg + \ldots + \int_{(q-1)\ell}^{q\ell} (M-q+1)mg + \int_{q\ell}^s (M-q)mg \right) \frac{\alpha(z)}{A(z)} dz$$

$$= mg \sum_{k=0}^{q-1} (M-k) \int_{k\ell}^{(k+1)\ell} \frac{\alpha(z)}{A(z)} dz + (M-q)mg \int_{q\ell}^s \frac{\alpha(z)}{A(z)} dz \qquad (8)$$

Equation (8) is a general equation applies for all fronds. We will investigate the properties of the equation and its applications in some special cases as well as comparison with observations.

## 3. RESULTS AND DISCUSSION

First we determine the bending angle formed by free end of the rod, which is located at $s = L$. The parameter $\Omega_f$ of the free end is

$$\Omega_f = [\tan\theta_0 + \sec\theta_0] \exp\left[ -g \int_0^L \frac{\alpha(z)}{A(z)} \left\{ \int_z^L \lambda(s') ds' \right\} dz - \int_0^L \frac{\alpha(z)}{A(z)} P\{z,L\} dz \right] \qquad (9)$$

and the angle formed by free end is $\theta_f = 2\tan^{-1}[(\Omega_f -1)/(\Omega_f +1)]$. For very long rods, the terms inside the exponential in Eq. (9) approaches negative infinity such that $\Omega_f \to 0$, implying $\theta_f \to 2\tan^{-1}(-1) \to -\pi/2$.

If the rod is mechanically very soft (large α), it is easily bent. If α is extremely large, the integral in the exponential function in equation (9) approaches infinity and we obtain a condition similar to the case of very long rods, i.e., $\theta_f \to -\pi/2$. This condition is likely shown by palm frond or banana leaves ends as shown in **Figure 1**. The mechanical strength of the palm frond decreases when moving toward the free end (toward the free end, the parameter α becomes larger).

If the rod is very strong, $\alpha \to 0$, so that the exponential term approcahes unity and based on Eq. (3b) we have $\theta = \theta_0$. These results indicate that the rod does not undergo any bending elsewhere.

Next, let us examine the location of bending peak where $\theta = 0$ or $\tan\theta + \sec\theta = 1$ so that $\Omega = 1$. If the rod is short enough or the fixed angle is large enough, the rod might have not a bending peak. In this case the bending angles are positive at all positions throught the rod. The bending peak occurs if there is $s_m$ such that, based on Eq. (3a),

$$\ln\left[\frac{1}{\tan\theta_0 + \sec\theta_0}\right] = -g\int_0^{s_m}\frac{\alpha(z)}{A(z)}\left\{\int_z^L \lambda(s')ds'\right\}dz - \int_0^{s_m}\frac{\alpha(z)}{A(z)}P\{z,L\}dz \tag{10}$$

It is clear that the bending peak location is strongly dependent on $\theta_0$. The maximum fixed angle (critical fixed angle) to produce the bending peak, $\theta_{0c}$, satisfies the following equation

$$\tan\theta_{0c} + \sec\theta_{0c} = \exp\left[g\int_0^L\frac{\alpha(z)}{A(z)}\left\{\int_z^L \lambda(s')ds'\right\}dz + \int_0^L\frac{\alpha(z)}{A(z)}P\{z,L\}dz\right]$$

giving rise to the solution $\theta_{0c} = 2\tan^{-1}[(\Omega_c^* -1)/(\Omega_c^* +1)]$, with

$$\Omega_c^* = \exp\left[ g\int_0^L \frac{\alpha(z)}{A(z)} \left\{\int_z^L \lambda(s')ds'\right\} dz + \int_0^L \frac{\alpha(z)}{A(z)} P\{z,L\}dz \right] \tag{11}$$

At this critical fixed angle, the bending peak is located at the free end ($s_m = L$). The bending peak is located at any positions along the rod when $\theta_0 < \theta_{0c}$. It is easily proven that the critical bending peak can be expressed as $\theta_{0c} = 2\tan^{-1}[\tanh(\Omega_c^*/2)]$.

For bare homogeneous rods, the mass density, cross section, and strengthness are independent of position so that $\lambda(s) = \lambda$, $A(s) = A$, and $\alpha(s) = \alpha$ for all $s$. The leaves are also not attached to the rod, or $m = 0$. These behaviors lead to

$$\int_0^s \frac{\alpha(z)}{A(z)} \left\{\int_z^L g\lambda(s')ds'\right\} dz = \frac{g\alpha\lambda}{A}\left(Ls - \frac{1}{2}s^2\right)$$

and the bending angle satisfies Eq (7) with $\Omega$ is given by

$$\Omega = [\tan\theta_0 + \sec\theta_0]\exp\left[-\kappa(2\xi - \xi^2)\right] \tag{12}$$

with $\kappa = g\alpha\lambda L^2/2A$, and $\xi = s/L$. Special case at the free end ($s = L$ or $\xi = 1$) we have $\Omega_f = [\tan\theta_0 + \sec\theta_0]\exp[-\kappa]$. Since $\kappa \propto L^2$, $\Omega_f$ decreases exponentially with square of the rod length. If $L$ is very large such that $\kappa \to \infty$, $\Omega_f \to 0$ and based on Eq. (7) we have $\theta_f \to 2\tan^{-1}(-1) = -\pi/2$.

If the rod is long enough or the fixed angle is not too large so that the bending peak occurs some where along the rod, the location of the bending peak, $s_m$, satisfies $\ln[1/(\tan\theta_0 + \sec\theta_0)] = -\kappa(2\xi_m - \xi_m^2)$ with $\xi_m = s_m/L$ to, giving rise to the following equation

$$\xi_m = 1 - \sqrt{1 - \frac{1}{\kappa}\ln[\tan\theta_0 + \sec\theta_0]} \tag{13}$$

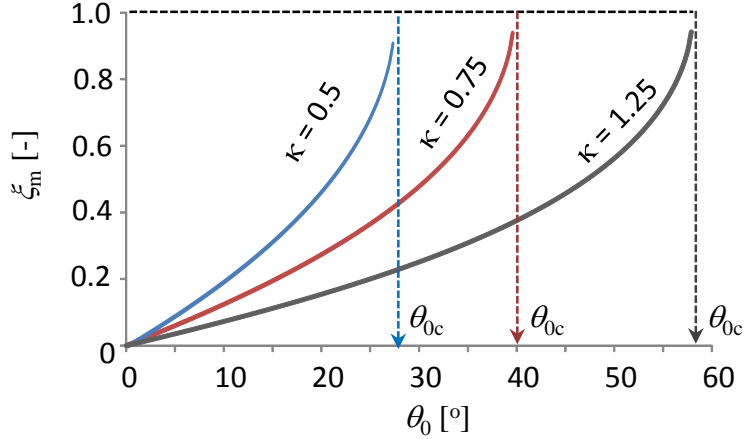

**Figure 4** Locations of bending peaks of homogeneous rod as a function of fixed angle at various values of κ = 0.5, 0.75, dan1.25. The right end of each curve represents $\theta_{0c}$.

It is clear from the equation (13) that for rods from the same material, thelocation of $\xi_m$ is highly dependent on the fixed angle and the rod length. **Figure 4** is illustrations of the calculated $\xi_m$ at various values of $\kappa = g\alpha\lambda L^2/2A$. Each curve has a right boundary, representing the critical fixed angle, $\theta_{0c}$. The critical fixed angle to produce a peak at the free end can be obtained from Eq. (13), satisfied by $1-(1/\kappa)\ln[\tan\theta_{0c}+\sec\theta_{0c}]=0$, or $\tan\theta_{0c}+\sec\theta_{0c}=\exp[\kappa]$. Based on Eq. (3b), (4), and (5), the solution for the critical fixed angle is

$$\theta_{0c} = 2\tan^{-1}\left[\frac{e^\kappa - 1}{e^\kappa + 1}\right]$$

$$= 2\tan^{-1}\left[\tanh\left(\frac{\kappa}{2}\right)\right] \tag{14}$$

Let us now inspect the behavior of $\xi_m$ around the critical fixed angle. We inspect the behavior at angle $\theta_0 = \theta_{0c} - \delta$, with $\delta \ll 1$ and $(1/\kappa)\ln[\tan\theta_{0c}+\sec\theta_{0c}]=1$. We approximate $\tan\theta_0 = \tan[\theta_{0c}-\delta] = (\sin\theta_{0c}\cos\delta - \cos\theta_{0c}\sin\delta)/(\cos\theta_{0c}\cos\delta + \sin\theta_{0c}\sin\delta) = (\tan\theta_{0c}-\delta)/(1+\delta\tan\theta_{0c}) \approx \tan\theta_{0c} - \delta\sec^2\theta_{0c}$. However, $\sec\theta_0 = \sec[\theta_{0c}-\delta] \approx \sec\theta_{0c}$. Therefore, Eq. (13) can be written as

$$1 - \xi_m \approx \sqrt{1 - \frac{1}{\kappa} \ln[\tan\theta_{0c} + \sec\theta_{0c} - \delta\tan\theta_{0c}]}$$

$$= \sqrt{1 - \frac{1}{\kappa} \ln\left[(\tan\theta_{0c} + \sec\theta_{0c})\left(1 - \frac{\tan\theta_{0c}}{\tan\theta_{0c} + \sec\theta_{0c}}\delta\right)\right]}$$

$$= \sqrt{1 - \frac{1}{\kappa} \ln[\tan\theta_{0c} + \sec\theta_{0c}] - \frac{1}{\kappa} \ln\left[1 - \frac{\tan\theta_{0c}}{\tan\theta_{0c} + \sec\theta_{0c}}\delta\right]}$$

$$= 1 - \sqrt{-\frac{1}{\kappa} \ln\left[1 - \frac{\tan\theta_{0c}}{\tan\theta_{0c} + \sec\theta_{0c}}\delta\right]}$$

$$\approx \sqrt{\frac{\kappa\tan\theta_{0c}}{\tan\theta_{0c} + \sec\theta_{0c}}\delta}$$

$$\propto (\theta_{0c} - \theta_0)^{1/2} \tag{15}$$

From this result we can state that the change of $1-\xi_m$ around the critical fixed angle is universal, i.e, exponential factor of ½.

To the contrary, when $\theta_0 \to 0$ we can approximate $\tan\theta_0 + \sec\theta_0 \approx \theta_0 + 1$ and $\ln[\tan\theta_0 + \sec\theta_0] \approx \theta_0$, Therefore, fron Eq. (13) we have the following approximation

$$\xi_m \approx 1 - \sqrt{1 - \frac{\theta_0}{\kappa}}$$

$$\approx 1 - \left(1 - \frac{\theta_0}{2\kappa}\right)$$

$$\propto \theta_0 \tag{16}$$

Thus, when the fixed angle approaches zero, the location of bending peak is proportional to the fixed angle.

***Palm Frond with Leaves***. For homogeneous rod attached by leaves, Eq (8) can be written as

$$\int_0^s \frac{\alpha(z)}{A(z)} P\{z,L\} dz = \frac{mg\alpha\ell}{A} \sum_{k=0}^{q-1}(M-k) + \frac{(M-q)mg\alpha}{A}(s-q\ell)$$

$$= \frac{mg\alpha}{2A}\{(q^2+q)\ell + 2(M-q)s\}$$

and

$$\Omega = [\tan\theta_0 + \sec\theta_0]\exp\left[-\frac{g\alpha\lambda}{A}\left(Ls - \frac{1}{2}s^2\right) - \frac{mg\alpha}{2A}\{(q^2+q)\ell + 2(M-q)s\}\right]$$

$$= [\tan\theta_0 + \sec\theta_0]\exp\left[-\kappa(2\xi - \xi^2) - \rho\kappa\left\{(q^2+q)\frac{\ell}{L} + 2(M-q)\xi\right\}\right] \quad (17)$$

where $\rho = m/\lambda L = m/m_b$, and $m_b = \lambda L$ is the frond mass.

An attractive behavior is obtained from Eq. (17) when $M \to \infty$ or $\ell \to 0$. If this condition is satisfied, $q = s/\ell$ is very large for most positions along the frond such that $q^2 + q \approx q^2$ and Eq. (16) can be approximated as

$$\Omega \approx [\tan\theta_0 + \sec\theta_0]\exp\left[-\kappa(2\xi - \xi^2) - \rho\kappa\left\{q^2\frac{\ell}{L} + 2(M-q)\xi\right\}\right]$$

$$= [\tan\theta_0 + \sec\theta_0]\exp\left[-\kappa(2\xi - \xi^2) - \frac{m}{\lambda L}\kappa\left\{\frac{s^2}{\ell^2}\frac{\ell}{L} + 2(M - \frac{s}{\ell})\xi\right\}\right]$$

$$= [\tan\theta_0 + \sec\theta_0]\exp\left[-(\kappa + \kappa')(2\xi - \xi^2)\right] \quad (18)$$

with $\kappa' = g\alpha\lambda' L^2/2A$ is contributed by leaves only. Equation (18) is exactly the same as the Eq. (12), only by changing the mass density (mass per unit length) to account also the masses of the leaves.

***Single force at free end***. Whe the non gravitational force is a single force at the freen end, by hanging a load of mass m we obtain $P\{z,L\} = mg$ for all $z$. Therefore, for homogeneous beam we obtain

$$\int_0^s \frac{\alpha(z)}{A(z)} P\{z,L\} dz = \frac{\alpha mgs}{A} \tag{19}$$

$$\Omega = [\tan\theta_0 + \sec\theta_0]\exp\left[-\frac{g\alpha\lambda}{A}\left(Ls - \frac{1}{2}s^2\right) - \frac{mg\alpha s}{A}\right] \tag{20}$$

At the free end we have

$$\Omega_f = [\tan\theta_0 + \sec\theta_0]\exp\left[-\frac{g\alpha L}{A}\left(\frac{\lambda L}{2} + m\right)\right]$$

$$= \exp\left[\ln(\tan\theta_0 + \sec\theta_0) - \frac{g\alpha L}{A}\left(\frac{\lambda L}{2} + m\right)\right] \tag{21}$$

and the angle made by the free end is

$$\theta_f = 2\tan^{-1}\left(\frac{\Omega_f - 1}{\Omega_f + 1}\right)$$

$$= 2\tan^{-1}\left(\tanh\left[\frac{\ln(\tan\theta_0 + \sec\theta_0)}{2} - \frac{g\alpha L}{2A}\left(\frac{\lambda L}{2} + m\right)\right]\right) \tag{22}$$

If the fixed angle is zero or the beam is positioned horizontally we have

$$\theta_f = 2\tan^{-1}\left(\tanh\left[-\frac{g\alpha L}{2A}\left(\frac{\lambda L}{2} + m\right)\right]\right) \tag{23}$$

If the load mass is much higher than the rod mass, we have $m >> \lambda L$ so that

$$\theta_f = 2\tan^{-1}\left(\tanh\left[-\frac{mg\alpha L}{2A}\right]\right) \tag{24}$$

***Proposal of time dependent frond.*** The length of the palm fronds changes with time. With increasing time the frond length increases until it reaches a saturation length after a certain period. The frond length initially increases rapidly and then slows down. With these properties, we may approximate the change of palm frond length with a simple equation $L(t) = L_\infty(1-e^{-\sigma t})$, where $L_\infty$ is the frond length as time approches infinity (saturation length) and $\sigma$ is a constant determining the rate length increase.

The palm frond mechanical strength depends on the position along the frond because of the different positions has been produced at different time. Towards the free end, the mechanical strength becomes weaker. However, if the frond has been produced at very long periode, the mechanical strength becomes saturated. Suppose a certain frond segment was producedat a time $\tau$. At $t>\tau$ the strength of the segment depends on $t-\tau$. Suppose the strangeness of the frond changes exponentially with time so that the strength of the rod, $K$, that has been produced at time $\tau$ satisfies $K(\tau,t) = K_\infty(1-e^{-\gamma(t-\tau)})$, with $K_\infty$ is the saturated mechanical strength and $\gamma$ is a constant determining the rate of strengness increase. It is clear that when $t\to\infty$ then $K\to K_\infty$ for all parts of the frond produced at arbitrary time $\tau$. At this time the frond has finished the process of growth and the physical propersies unchange anymore.

The frond segment at distance $s$ from the fixed end has been produced at $\tau$ satisfying $s = L(\tau) = L_\infty(1-e^{-\sigma\tau})$ or $e^{-\sigma\tau} = 1-s/L_\infty$. At $t>\tau$, the frond length is $L(t) = L_\infty(1-e^{-\sigma t})$ or $e^{-\sigma t} = 1-L(t)/L_\infty$. From the above equations we obtain $e^{-\sigma(t-\tau)} = [1-L(t)/L_\infty]/[1-s/L_\infty]$. Thus the strength of the frond as a function of position can be written as

$$K(s) = K_\infty \left\{1 - \left[\frac{L_\infty - L}{L_\infty - s}\right]^{\gamma/\sigma}\right\} \qquad (25)$$

There are two parameters determining the mechanical strength of the rod in various positions, namely the gowth parameter, $\sigma$, and the parameter mechanical strength increasing, $\gamma$. By remembering that $\alpha$ is inversely proportional to the mechanical strength of the frond we obtain the equation dependence of $\alpha$ on the position along the frond as follows

$$\alpha(s) = \frac{\alpha_\infty}{1 - \left[\dfrac{L_\infty - L}{L_\infty - s}\right]^{\gamma/\sigma}} \qquad (26)$$

**Figure 5(a)** is an example of dependence of bending proportional constant on distance from the frond fixed end calculated using Eq. (26) at different values of $L/L_\infty$. For illustration we used $\gamma/\sigma = 0.5$. Boths axes were normalized to saturated values, i.e., the corrsponding values at $t \rightarrow \infty$. The figure shows that up to distance 65% fron the fixed end, the proportional constant changes slightly with distance. Significant change was observed form distances between 65% to 100% of the frond length. This result means that drastic change in the frond bending occurs at last 35% portion of the frond.

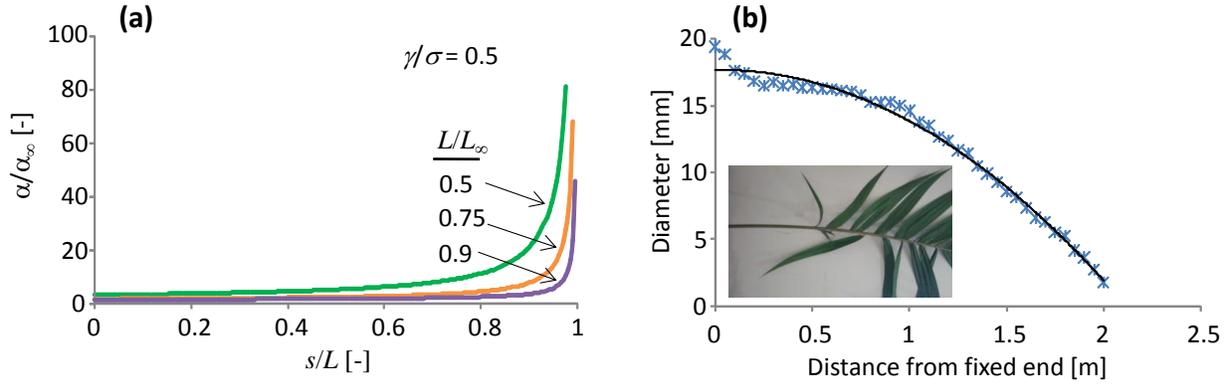

**Figure 5(a)** Ilustration of dependence of proportional constant on distance from the frond fixed end (normalized to frond length) at different values of $L/L_\infty$. For ilustration we used $\gamma/\sigma = 0.5$. **(b)** Depedence of frond diameter on distance from the fixed end. Inset is the picture of frond that has been measured. Symbols are measured data and curve is the fitting result.

Once we hypothesized function $A(s)$, $\alpha(s)$, and $\lambda(s)$ for the palm frond, Eq. (9) becomes

$$\ln\left[\frac{\tan\theta + \sec\theta}{\tan\theta_0 + \sec\theta_0}\right] = -g\alpha_\infty \mu \int_0^s \left\{ \frac{\int_z^L A(s')ds'}{A(z)\left(1 - [(L_\infty - L)/(L_\infty - z)]^{\gamma/\sigma}\right)} \right\} dz$$

$$-\alpha_\infty \int_0^s \left\{ \frac{P\{z,L\}}{A(z)\left(1-\left[(L_\infty - L)/(L_\infty - z)\right]^{\gamma/\sigma}\right)} \right\} dz \qquad (27)$$

Indeed, Eq. (27) can be solved for any function $A(z)$ or $P\{z,L\}$ if all parameters are known.

We measured example of a palm frond diameter as function of posision from the fixed end. **Figure 5(b)** shows the mesurement data of the frond diameter (symbols) and the inset picture is the frond has been measured. The measurement data can be best fit using equation $d(s) = -4.065s^2 + 0.246s + 17.67$ [mm] with $s$ in meter ($R^2 = 0.988$). This function is represented by curve in **Figure 5(b).** Based on this result, we assume the dependence of frond radius on position is given by $r(s) = a_1 s^2 + a_2 s + R_0$, with $a_1$, $a_2$, and $R_0$ depend on time. Different fronds may have different function for $r(s)$.

**Comparison with Measurement Data**

To obtain the bending profile of the rod, we used the following iteration equation, $x_0 = y_0 = 0$, $x_1 = a\cos\theta_0$, $y_1 = a\sin\theta_0$, $x_2 = x_1 + a\cos\theta_1$, and $y_2 = y_1 + a\sin\theta_1$, or in general

$$x_n = x_{n-1} + \cos\theta_{n-1} \quad \text{and} \quad y_n = y_{n-1} + a\sin\theta_{n-1} \qquad (28)$$

From these coordinates, we can draw the bending profile of the rod at various fixed angles.

*Effects of fixed angle and rod length on bending.* **Figure 6(a)** is examples of bending profiles at various fixed angles: 70°, 45°, and 28°. Symbols are measurement data and curves are calculation results. For calculation we used $\kappa = 1.053$, selected so that all curves better fit the measured data at all fixed angles. The measurement data were obtained using an iron ruler of 1 m length, 3 cm width, 1 mm thick, and 218 g mass (mass density $\lambda = 0.218$ kg/m). We see the simulation results are in accordance with measurement data. From the fitting results we estimated $\alpha$ for the ruler as $\alpha = 2A \times 1.053/g\lambda L^2 = 2.95 \times 10^{-5}$ s²/kg.

**Figure 6(b)** is bending profile having rods of different lengths. Symbols are measurement data and curves are calculation results. For ruller of length 1 meter we used $\kappa = $ 1.053. Since this parameter is proportional to the square of the length then for ruller of lengths 0.75 m and 0.5 m we used $\kappa = 0.592$ and $\kappa = 0.263$, respectively. In all experiment the rullers were tilted at the same angle of $28^o$. It also appears here that the simulation results fairly fit the measurement data.

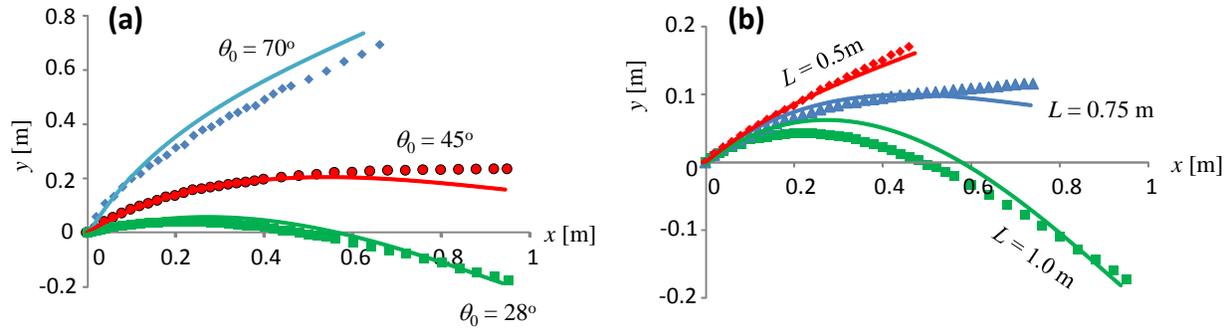

**Figure 6(a)** Bending profiles of rullers at different fixed angles: $70^o$, $45^o$, and $28^o$. The symbols are measurement data and the curves are the calculations results using equations (4), (5), and (28). For all calculations we used $\kappa = 1.053$. **(b)** Bending profiles of rullers of different lengths. For ruller of length 1 m, 0.75 m, dan 0.5 we used $\kappa = 1.053$, 0.592, and 0.263, respectively. This parameter is inversely proportional to the square of the length. All rullers were tilted at the same angle of $28^o$.

*Effect of leaves attached to frond on bending*. **Figure 7** is bending profiles of a bare rod (diamond symbols), a rod attached by PVC pieces of the same spacing: (squares) for $M = 5$ and (triangles) for $M = 9$. Symbols are the measurement data and the curves are the calculation results. We used the same iron ruller. In the experiment we attached pieces of PVC with a length of 20 cm and a mass of each piece is $m = 7.7$ g. In the calculation we kept using $\kappa = 1.053$. All rods were tilted at a fixed angle $56^o$. It also appears that the calculation result is quite close to the measurement results.

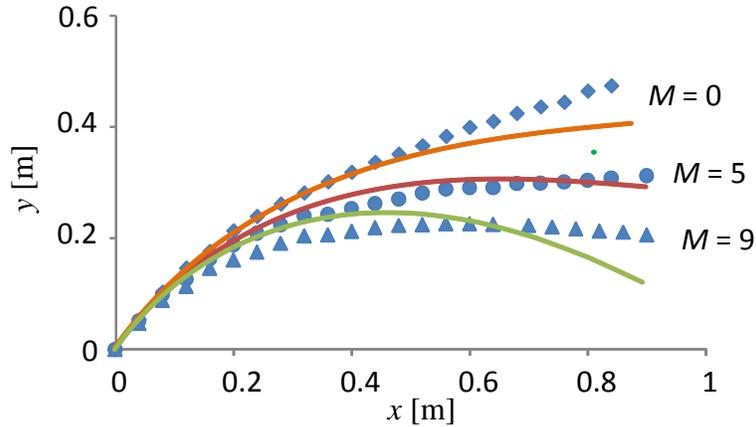

**Figure 7 B**ending profile of bare rod (diamond symbols), a rod attached with PVC pieces placed at the same spacing: (squares) for $M = 5$ and (triangles) for $M = 9$. Symbols are the measurement results and curves are the calculation results. We used the same iron ruller.

To the end, we have succedded to derive a new and general equation applied to explain bending of arbitrary rods and well as palm frond in a very simple way. No iteration procedure is required to find the bending profile. It seems that several bending problems that usually obtained by long numerical calculation can be simplified.

**CONCLUSION**

Equations (4) and (5) are the new and general equation we have derived for explain bending of arbitrary rodd with arbitraty length, cross section, mass density, strengthness, bending angle, etc in a very simple way. No complex mathematical formulation or numerical procedure is required. The equation can be used also to explain bending of palm fronds, banana leaves or other spahes in nature regarding to bending dua to gravity. The equation can explain well experimental data, either of bare rods or frond like rods.

**AKCNOWLEDGEMENT**

This work was financially supported by the Ministry of Research and Higher Education, Republic of Indonesia (Grant No. 310y/I1.C01/PL/2015).